\documentstyle[amsmath,mathrsfs,a4,11pt]{article}

\date{}

\begin{document}

\title{{\bf Gravitational wave amplification of seed magnetic
fields}}

\author{Christos G. Tsagas$^1$, Peter K. S. Dunsby$^{1,2}$ and
Mattias Marklund$^3$\\{\small $^1$Department of Mathematics and
Applied Mathematics, University of Cape Town, Rondebosch 7701,
South Africa}\\{\small $^2$South African Astronomical Observatory,
Observatory 7925, Cape Town, South Africa}\\
{\small $^3$~Department of Electromagnetics, Chalmers University
of Technology, SE--412 96 G\"oteborg, Sweden}}

\maketitle \maketitle

\begin{abstract}
We discuss how gravitational waves could amplify seed magnetic
fields to strengths capable of supporting the galactic dynamo. We
consider the interaction of a weak magnetic field with gravity
wave distortions in almost FRW cosmologies and find that the
magnitude of the original field is amplified proportionally to the
wave induced shear anisotropy and, crucially, proportionally to
the square of the field's initial scale. The latter makes our
mechanism particularly efficient when operating on superhorizon
sized magnetic fields, like those produced during inflation. In
that case, the achieved amplification can easily boost magnetic
strengths, which may still lie relatively close to the galactic
dynamo lower limits, well within the currently accepted range.\\\\
PACS number(s): 98.80.Hw, 04.30.-w, 98.80.Cq
\end{abstract}

\maketitle

\section{Introduction}
Large scale magnetic fields, with strengths between $10^{-7}$ and
$10^{-5}~G$, have been repeatedly observed in spiral and disc
galaxies, in galaxy clusters as well as in high redshift
condensations~\cite{K}. Despite their established widespread
presence, however, the origin of cosmic magnetic fields remains a
mystery and is still a matter of debate~\cite{GR}. Over the years,
a number of possible solutions has been proposed, ranging from
eddies and density fluctuations in the early plasma to
cosmological phase-transitions, inflationary and superstring
inspired scenarios (see~\cite{H,D} for a representative list).
Historically, studies of magnetogenesis were motivated by the need
to explain the origin of the large-scale galactic fields. Typical
spiral galaxies have magnetic fields of the order of a few $\mu$G
coherent over the plane of their disc. The structure of these
fields, particularly those in spiral galaxies, strongly suggests
that they have been generated and sustained by a dynamo
mechanism~\cite{P}. Although the efficiency of the mechanism has
been critisised, it is generally believed that galactic dynamos
can substantially amplify preexisting weak magnetic fields by
combining the turbulent motion of the ionised gas with the
differential rotation of the galaxy. The origin of the required
seed fields, however, is still elusive. They could be the result
of local astrophysical processes, such as buttery and vorticity
effects, or the remnants of a large scale primordial magnetic
field. Provided that the nonlinear dynamo amplification is
efficient, the seeds can be as low as $\sim10^{-23}~G$ at
present~\cite{Ku}. For a spatially flat universe dominated by
``dark-energy'', namely by a cosmological constant or
quintessence, the aforementioned lower limit is further relaxed
down to $\sim10^{-30}~G$~\cite{DLT}. In the absence of a dynamo
mechanism, however, magnetic seeds of the order of $10^{-12}~G$,
or even $10^{-8}~G$, are required. The coherence scale of the seed
field is an additional issue. Typically, galactic dynamos require
a minimum coherence length comparable to the dimensions of the
largest turbulent eddy, which is of the order of 100~pc, to
guarantee the stability of the amplification process~\cite{KA}.

The attractiveness of primordial magnetic fields lies in the fact
that they can readily explain both the fields seen in nearby
galaxies as well as those detected in galaxy clusters and highly
redshifted condensations. There have been numerous attempts to
generate early, pre-recombination, magnetic fields by exploiting
the different out-of-equilibrium epochs that are believed to have
taken place between the end of the inflationary era and
decoupling. In all these scenarios, however, the causal nature of
the generating mechanisms means that the coherence scale of the
induced seed fields is unacceptably small. A process known as
``inverse cascading'' can provide a solution to the incoherence
problem by transferring magnetic energy to increasingly larger
scales~\cite{C}. This mechanism, however, requires a considerable
amount of net helicity in the cosmic fluid and therefore scenarios
based on inverse cascade are still treated as rather speculative.
Inflation has long been suggested as a solution to the causality
problem, since it naturally achieves correlations on superhorizon
scales. Nevertheless, the conformal invariance of electromagnetism
implies that any magnetic field present during the inflationary
regime will be strongly diluted by the rapid expansion of the
universe. One can get around this obstacle by breaking the
conformal invariance of the gauge fields involved~\cite{Do,TW}.
There are more than one ways of doing that, which explains the
variety of the proposed mechanisms in the literature. For example,
there have been attempts to create magnetic fields by coupling the
photon to a scalar field either during inflation or in the
subsequent era of preheating~\cite{CKM}. These proposals have
since been criticised in~\cite{GS}. Other authors have advocated
the breakdown of Lorentz invariance either in the context of
string theory and non-commutative varying speed of light theories,
or due to the dynamics of large extra dimensions~\cite{BM}. The
success of these proposals, however, is usually achieved at the
expense of simplicity.

An inflation based mechanism that produces large scale magnetic
fields with strengths that could support the dynamo amplification
was recently proposed in~\cite{D}. One of the attractive aspects
of the approach, which exploits the natural coupling between the
Z-boson and the gravitational background during inflation, is that
it operates within the standard model. The magnitude of the
generated magnetic field, however, corresponds to $\sim10^{-30}~G$
on a collapsed scale of approximately 100~pc today, which is the
minimum magnetic strength required for the nonlinear galactic
dynamo to operate (in a dark-energy dominated universe). Even when
taking into account additional amplification (of up to 5 orders of
magnitude) during reheating, the achieved magnetic strength
remains rather uncomfortably close to the dynamo margin.
Nevertheless, the mechanism proposed in~\cite{D} is very
promising, as it clearly shows that magnetic fields which survive
inflation are not necessarily as weak as previously anticipated.

A common feature in all inflationary models is the production of
gravitational radiation with wavelengths extending from about 1~km
to $\sim3000$~Mpc today. In fact, a relic gravity wave spectrum is
perhaps the only direct signature of inflation that may still be
observable today. The coupling of these inflation produced gravity
waves with large scale magnetic fields, which may also be present
soon after inflation, could considerably affect the latter. In the
present article we try to address this issue within the framework
of standard general relativity. We consider the interaction
between the aforementioned two sources in a spatially flat FRW
cosmology during the radiation and the dust eras. Our results show
that, in the presence of gravitational radiation, the magnitude of
the magnetic field is amplified proportionally to the shear
distortion caused by the propagating waves. Crucially, however,
the gravitational boost is also proportional to the square of the
field's original scale. This immediately suggests that the
mechanism presented here could lead to significant amplification
when dealing with large scale magnetic fields. Indeed, when
applied to fields of roughly $10^{-34}~G$ spanning a comoving
scale of about 10~kpc today, like those produced in~\cite{D}, our
mechanism leads to an amplification of up to 14 orders of
magnitude. The size of the boost can easily bring these magnetic
fields well within the galactic dynamo requirements, without the
need for extra amplification during reheating. In fact, the
enhancement is so effective that it can bring the field within the
dynamo limits even within conventional cosmological models which
are not dark-energy dominated. The latter task is more easily
achieved when the extra strengthening of the field, due to the
adiabatic collapse of the protogalaxy, is also taken into account.

\section{The model}
Consider a spatially flat FRW universe filled with a barotropic
perfect fluid and allow for the presence of a weak (test) magnetic
field $B_a$. For a fundamental observer, moving with 4-velocity
$u_a$ ($u_au^a=-1$), the stress-energy tensor of the magnetic
field reads~\cite{E,TB}
\begin{equation}
{\cal T}_{ab}={\textstyle{1\over2}}B^2u_au_b+
{\textstyle{1\over6}}B^2h_{ab}+ \Pi_{ab}\,, \label{cTab}
\end{equation}
where $h_{ab}=g_{ab}+u_au_b$ projects orthogonal to $u_a$ and
$g_{ab}$ is the spacetime metric. Note that in the absence of
vorticity $h_{ab}$ is the metric of the hypersurfaces orthogonal
to $u^a$. Also, ${\rm D}_a=h_a{}^b\nabla_b$ defines the covariant
derivative operator orthogonal to $u_a$. The scalar $B^2=B_aB^a$
measures the energy density and the isotropic pressure of the
field, while $\Pi_{ab}=-B_{\langle a}B_{b\rangle}$ describes the
anisotropic magnetic pressure.\footnote{Angled brackets indicate
the projected, symmetric and trace-free part of spacelike tensors,
while square brackets indicate their antisymmetric part.
Furthermore, we will also use $\text{curl}B_a =
\varepsilon_{abc}{\rm D}^bB^c$, where $\varepsilon_{abc}$ is the
projected permutation tensor (see~\cite{E}).} The weakness of the
field implies that its energy density, its anisotropic stresses
and its spatial gradients have negligible contribution to the
background dynamics. In other words, $B^2\ll\rho$ and
$\Pi_{ab}\simeq0\simeq{\rm D}_bB_a$ to zero order. In this limit,
the key background equations are
\begin{eqnarray}
\kappa\rho- {\textstyle{1\over3}}\Theta^2&=&0\,, \label{Fe}\\
\dot{\rho}+ (1+w)\Theta\rho&=&0\,, \label{ce}\\ \dot{B}_a+
{\textstyle{2\over3}}\Theta B_a&=&0\,, \label{mi}
\end{eqnarray}
where $\rho$ is the fluid energy density, $\Theta=3\dot{a}/a=3H$
is the expansion scalar (with $a$ and $H$ representing the scale
factor and the Hubble parameter respectively) and $w=p/\rho$
(where $p$ is isotropic pressure of the fluid). Equations
(\ref{Fe}), (\ref{ce}) and (\ref{mi}) are respectively the
Friedman equation, the equation of continuity and the magnetic
induction equation. For a fully covariant description of
electromagnetic fields in an expanding universe the reader is
referred to~\cite{TB}. Finally, we should point out that
throughout the paper we employ natural units with $c=1=\hbar$ and
$\kappa=8\pi G=m_{Pl}^{-2}$.

We perturb the aforementioned FRW background by allowing for weak
gravitational waves, and employ the covariant and gauge-invariant
perturbation formalism (see~\cite{E}-\cite{MDB} for details),
which guarantees that our results are free from gauge ambiguities.
Covariantly, gravity waves are monitored via the electric
($E_{ab}$) and the magnetic ($H_{ab}$) Weyl components, which are
the symmetric and trace-free tensors that describe the long-range
gravitational field~\cite{DBE}. In the magnetic presence, one
isolates the linear tensor perturbations, namely switches all the
scalar and the vector modes off, by imposing the following
conditions~\cite{MTU}
\begin{equation}
{\rm D}_aB^2=0=\varepsilon_{abc}B^b{\rm curl}B^c\,, \label{cnt1}
\end{equation}
in addition to the standard constraints
\begin{equation}
\omega_a=0={\rm D}_a\rho={\rm D}_ap\,,  \label{cnt2}
\end{equation}
associated with the pure perfect fluid cosmologies~\cite{DBE}.
Together, constraints (\ref{cnt1}) and (\ref{cnt2}) guarantee that
all traceless tensors are transverse to first order and therefore
isolate the pure tensor (i.e.~the transverse traceless)
modes~\cite{MTU}. Note that condition (\ref{cnt1}) means that we
are dealing with a force-free magnetic field. This in turn implies
that any spatial currents that might have been present are also
switched off.

\section{The interaction}
Having set the mathematical framework, we now proceed to look into
the linear interaction between magnetic fields and propagating
gravitational radiation. To first order, the magnetic field
evolution in the presence of gravity wave perturbations is
governed by the system\footnote{During linearisation the
perturbative order of the various quantities is determined by
their background value. Quantities with a non-zero unperturbed
value have zero perturbative order, while those that vanish in the
background are of order one.}
\begin{eqnarray}
\ddot{B}_a- {\rm D}^2B_a+ {\textstyle{5\over3}}\Theta\dot{B}_a+
{\textstyle{1\over3}}\Theta^2(1-w)B_a&=&
{\textstyle{4\over3}}\Theta\tilde{B}^b\sigma_{ab}+
2\tilde{B}^b\dot{\sigma}_{ab}\,, \label{ddotB}\\
\ddot{\sigma}_{ab}- {\rm D}^2\sigma_{ab}+
{\textstyle{5\over3}}\Theta\dot{\sigma}_{ab}+
{\textstyle{1\over6}}\Theta^2(1-3w)\sigma_{ab}&=&0\,.
\label{ddotsigma}
\end{eqnarray}
Here, $\tilde{B}_a$ is the original background magnetic field, and
$B_a$ is the perturbed field that results from the coupling of
$\tilde{B}$ with gravity wave distortions. Thus, to first order,
only the background magnetic field contributes to the right-hand
side of (\ref{ddotB}). Note that we have ignored the magnetic
contribution to the right-hand side of Eq.~(\ref{ddotsigma}),
given that the linear evolution of the shear is effectively immune
to magnetic effects. Indeed, the magnetic presence induces
decaying shear modes, which lie in between the standard
ones~\cite{MTU}. Also, in deriving Eq.~(\ref{ddotsigma}) we have
employed the linear propagation equation of the shear
\begin{equation}
\dot{\sigma}_{ab}=-{\textstyle{2\over3}}\Theta\sigma_{ab}+
{\textstyle{1\over2}}\kappa\Pi_{ab}- E_{ab}\,,  \label{dotsigma}
\end{equation}
which allowed us to express the electric Weyl tensor in terms of
$\sigma_{ab}$ and $\Pi_{ab}$. Finally, on using the linear
relation $H_{ab}={\rm curl}\sigma_{ab}$, we have replaced the
magnetic Weyl tensor with the shear. We should point out that the
gravity waves are the driving force behind the
``gravito-magnetic'' fluctuations in Eq.~(\ref{ddotB}). In
particular, one can explicitly show that gravitational waves
trigger fluctuations in an otherwise homogeneous background
magnetic field distribution via the magnetic part of the Weyl
tensor (i.e.~via the shear to first approximation)~\cite{TB}.

According to Eq.~(\ref{mi}), the background magnetic field evolves
as $\tilde{B}_a=\tilde{B}_a^0(a_0/a)^2$, where
$\dot{\tilde{B}}_a^0=0$ and the zero suffix corresponds to a given
initial time. Assuming that $\tilde{B}_a^0=\tilde{B}_{({\rm
n})}^0A_a^{({\rm n})}$, with $\dot{A}_a^{({\rm n})}=0$ and ${\rm
D}^2A_a^{({\rm n})}=-({\rm n}^2/a^2)A_a^{({\rm n})}$, enables us
to define a physical coherence scale $\lambda_{\tilde{B}}=2\pi
a/{\rm n}$ for $\tilde{B}_a$~\cite{MDB}. Then, introducing the
tensor harmonics $Q^{({\rm k})}_{ab}$, with $\dot{Q}_{ab}^{({\rm
k})}=0$ and ${\rm D}^2Q_{ab}^{({\rm k})}=-({\rm
k}^2/a^2)Q_{ab}^{({\rm k})}$, we adopt the decomposition
$\sigma_{ab}=\sigma_{({\rm k})}Q^{({\rm k})}_{ab}$ for the shear.
Note that both $A_a^{({\rm n})}$ and $Q_{ab}^{({\rm k})}$ are
dimensionless and unitary. On these grounds, Eqs.~(\ref{ddotB}) and
(\ref{ddotsigma}) decompose into the following system of ordinary
differential equations:
\begin{eqnarray}
\ddot{B}_{({\ell})}+
{\textstyle{5\over3}}\Theta\dot{B}_{({\ell})}+
\left[{\textstyle{1\over3}}(1-w)\Theta^2
+\frac{\ell^2}{a^2}\right]B_{({\ell})}&=&
2\left(\dot{\sigma}_{({\rm k})}+
{\textstyle{2\over3}}\Theta\sigma_{({\rm
k})}\right)\tilde{B}_0^{({\rm n})}\left(\frac{a_0}{a}\right)^2\,,
\label{ddotB1}\\ \ddot{\sigma}_{({\rm k})}+
{\textstyle{5\over3}}\Theta\dot{\sigma}_{({\rm k})}+
\left[{\textstyle{1\over6}}(1-3w)\Theta^2+\frac{{\rm
k}^2}{a^2}\right]{\sigma}_{({\rm k})}&=&0\,,  \label{ddotsigma1}
\end{eqnarray}
with $B_a=B_{({\ell})}B^{({\ell})}_a$. In deriving
Eq.~(\ref{ddotB1}) we have used the vector harmonics
$B^{({\ell})}_a$, defined by
\begin{eqnarray}
B^{({\ell})}_a&=&Q_{ab}^{({\rm k})}A_{({\rm n})}^b\,, \label{Bl}
\end{eqnarray}
which are also dimensionless with an order of unit magnitude. In
the above, $\ell$ and ${\rm n}$ are the comoving wavenumbers of the
perturbed and the background magnetic fields, respectively, while
${\rm k}$ is the comoving wavenumber of the gravitational wave.

It is rather straightforward to show that $B^{({\ell})}_a$
satisfies the standard vector-harmonic requirements. Indeed,
starting from definition (\ref{Bl}) we find that
$\dot{B}^{({\ell})}_a=0$, and then ${\rm
D}^2B^{({\ell})}_a=-(\ell^2/a^2)B^{({\ell})}_a$, where
$\ell^2=({\rm k}_a+{\rm n}_a)({\rm k}^a+{\rm n}^a)$. Clearly, the
wavenumber $\ell$ provides a measure of the scale of the
``induced'' magnetic field, which depends on the scale of the
background field and on the wavelength of the inducing
gravitational radiation. For ${\rm n}^a{\rm k}_a=0$, we arrive at
the simple expression
\begin{equation}
\lambda_B=\lambda_{\tilde{B}}[1+(\lambda_{\tilde{B}}/\lambda_{GW})^2]^{-1/2}\,,
\label{lambdaB}
\end{equation}
for the scale of the perturbed magnetic field. Note that
$\lambda_{\tilde{B}}=2\pi a/{\rm n}$, $\lambda_{GW}=2\pi a/{\rm
k}$ and $\lambda_{B}=2\pi a/\ell$ are the physical wavelengths of
the background magnetic field, the gravitational waves and the
perturbed magnetic field respectively. According to the above,
$\lambda_B\leq\lambda_{\tilde{B}}$. In particular,
$\lambda_B\sim\lambda_{\tilde{B}}$ as long as
$\lambda_{GW}\geq\lambda_{\tilde{B}}$, and
$\lambda_B<\lambda_{\tilde{B}}$ for
$\lambda_{GW}<\lambda_{\tilde{B}}$. Here, we will concentrate on
large scale magnetic fields and consider their interaction with
gravitational radiation of similar or larger wavelength.

Our final step is to introduce the expansion normalized
dimensionless variables
${\mathscr{B}}_{({\ell})}=\kappa^{1/2}B_{({\ell})}/\Theta$ and
$\Sigma_{({\rm
k})}=\sigma_{({\rm k})}/\Theta$, and rewrite Eqs.~(\ref{ddotB1}),
(\ref{ddotsigma1}) with respect to conformal time $\eta$ (defined
by $\dot{\eta}=a^{-1}$) as follows
\begin{eqnarray}
{\mathscr{B}}''_{({\ell})} +
(1-3w)\left(\frac{a'}{a}\right){\mathscr{B}}'_{({\ell})} -
\left[{\textstyle{3\over2}}(1-3w)w\left(\frac{a'}{a}\right)^2
-\ell^2\right]{\mathscr{B}}_{({\ell})}&=&\nonumber\\
2\kappa^{1/2}a\left[\Sigma'_{({\rm k})}
+{\textstyle{1\over2}}(1-3w)\left(\frac{a'}{a}\right)\Sigma_{({\rm
k})}\right]\tilde{B}_0^{(n)}\left(\frac{a_0}{a}\right)^2\,,
\label{ddotcB}\\ \Sigma''_{({\rm k})}+
(1-3w)\left(\frac{a'}{a}\right)\Sigma'_{({\rm k})}-
\left[{\textstyle{3\over2}}\left[1+(2-3w)w\right]
\left(\frac{a'}{a}\right)^2 -{\rm k}^2\right]\Sigma_{({\rm
k})}&=&0\,. \label{ddotSigma}
\end{eqnarray}
where the prime indicates differentiation with respect to $\eta$.

\section{The effect}
Let us consider the effects described by Eqs.~(\ref{ddotcB}) and
(\ref{ddotSigma}) during different epochs in the lifetime of the
universe. For $p=0$, which is the equation of state during dust
domination and also the effective equation of state at reheating,
$w=0$ and $a'/a=2/\eta$. Then, Eqs.~(\ref{ddotcB}) and
(\ref{ddotSigma}) reduce to
\begin{eqnarray}
{\mathscr{B}}''_{({\ell})}+ \frac{2}{\eta}{\mathscr{B}}'_{({\ell})}+
\ell^2{\mathscr{B}}_{({\ell})}&=&
\frac{8\alpha_1}{\eta^2}\left(\Sigma'_{({\rm
k})}+\frac{1}{\eta}\Sigma_{({\rm k})}\right)\,, \label{ddcB}\\
\Sigma''_{({\rm k})}+ \frac{2}{\eta}\Sigma'_{({\rm k})}-
\left(\frac{6}{\eta^2}-{\rm k}^2\right)\Sigma_{({\rm k})}&=&0\,,
\label{ddSigma}
\end{eqnarray}
where $\alpha_1=\kappa^{1/2}\tilde{B}_0^{({\rm n})}/a_0H_0^2$
depends entirely on the initial conditions. In the ${\rm
k}\eta\ll1$ limit, namely for long wavelength gravity waves, the
solution of Eq.~(\ref{ddSigma}) has a dominant mode given by
\begin{equation}
\Sigma^{({\rm k})}=\Sigma^{({\rm k})}(\eta)=\Sigma_0^{({\rm
k})}\left(\frac{\eta}{\eta_0}\right)^2\,, \label{Sigma}
\end{equation}
with $\eta_0^2=4/a_0^2H_0^2$ (see also~\cite{DBE}). Substituting
this result into Eq.~(\ref{ddcB}) we arrive at
\begin{equation}
{\mathscr{B}}^{({\rm n})}={\mathscr{B}}^{({\rm n})}(\eta)= {\mathscr{B}}_0^{({\rm
n})}\left(\frac{\eta_0}{\eta}\right)\left[\frac{{\rm cos}({\rm
n}\eta)}{{\rm cos}({\rm n}\eta_0)}\right]+
\frac{6\beta_1}{\eta}\,, \label{dcB}
\end{equation}
where $\beta_1=\kappa^{1/2}a_0\tilde{B}_0^{({\rm
n})}\Sigma_0^{({\rm k})}/{\rm n}^2$. The first term in the
right-hand side of the above solves the homogeneous equation and
the second conveys the gravity wave effects. On large scales ${\rm
n}\eta\ll1$ and (\ref{dcB}) reduces to
\begin{equation}
{\mathscr{B}}^{({\rm n})}= {\mathscr{B}}_0^{({\rm
n})}\left(\frac{\eta_0}{\eta}\right)+ \frac{6\beta_1}{\eta}\,.
\label{dcB1}
\end{equation}
Note that for large scale gravity waves, with
$\lambda_{GW}\sim\lambda_{\tilde{B}}$, we have
$\lambda_{B}\sim\lambda_{\tilde{B}}$ and therefore $\ell\sim{\rm
n}$ (see Eq.~(\ref{lambdaB})). In other words, the wavelength of
the perturbed magnetic field effectively coincides with that of
the background field. Given that ${\mathscr{B}}^{({\rm
n})}=\kappa^{1/2}B^{({\rm n})}/\Theta$ and $B_0^{({\rm
n})}=\tilde{B}_0^{({\rm n})}$ by definition, the above is recast
into the following expression for the magnetic field evolution
\begin{equation}
B^{({\rm n})}= \tilde{B}_0^{({\rm n})}\left[1
+9\left(\frac{\lambda_{\tilde{B}}}{\lambda_H}\right)_0^2\Sigma_0\right]
\left(\frac{a_0}{a}\right)^2\,. \label{dB}
\end{equation}
Here $(\lambda_{\tilde{B}})_0=a_0/{\rm n}$ and
$(\lambda_H)_0=1/H_0$ are the scale of the background
magnetic field and the horizon size at a given initial time,
respectively.
According to (\ref{dB}), the interaction of the background
magnetic field with gravity wave distortions can lead to
substantial increase of the field if
$9(\lambda_{\tilde{B}}/\lambda_H)_0^2\Sigma_0\gg1$.

Similar effects are also observed during the radiation era. When
relativistic species dominate the energy density of the universe
$w=1/3$ and $a'/a=1/\eta$. Then Eqs.~(\ref{ddotcB}),
(\ref{ddotSigma}) become
\begin{eqnarray}
{\mathscr{B}}''_{({\ell})}+ \ell^2{\mathscr{B}}_{({\ell})}&=&
\frac{2\alpha_2}{\eta}\Sigma'_{({\rm
k})}\,, \label{rdcB}\\
\Sigma''_{({\rm k})}- \left(\frac{2}{\eta^2}-{\rm
k}^2\right)\Sigma_{({\rm k})}&=&0\,, \label{rdSigma}
\end{eqnarray}
where $\alpha_2=\kappa^{1/2}\tilde{B}_0^{({\rm n})}/H_0$. For
${\rm k}\eta\ll1$ the solution of (\ref{rdSigma}) has a dominant
mode given by $\Sigma^{({\rm k})}=\Sigma_0^{({\rm
k})}\eta^2/\eta_0^2$, which when substituted into Eq.~(\ref{rdcB})
leads to the solution
\begin{equation}
{\mathscr{B}}^{({\rm n})}={\mathscr{B}}^{({\rm n})}(\eta)= {\mathscr{B}}_0^{({\rm
n})}\left[\frac{\cos({\rm n}\eta)}{\cos({\rm n}\eta_0)}\right]+
4\beta_2\,, \label{rcB}
\end{equation}
with $\beta_2=\kappa^{1/2}a_0^2H_0\tilde{B}_0^{({\rm
n})}\Sigma_0^{({\rm k})}/{\rm n}^2$. Confining, as before, to
large scale fields only (i.e.~${\rm n}\eta\ll1$) we arrive at the
following evolution law for the perturbed magnetic field
\begin{equation}
B^{({\rm n})}= \tilde{B}_0^{({\rm n})}\left[1
+12\left(\frac{\lambda_{\tilde{B}}}{\lambda_H}\right)_0^2\Sigma_0\right]
\left(\frac{a_0}{a}\right)^2\,. \label{rB}
\end{equation}

Results (\ref{dB}), (\ref{rB}) show that the presence of gravity
waves does not alter the radiation like evolution of the magnetic
field. Thus, dividing either (\ref{dB}) or (\ref{rB}) by the
energy density of the background radiation field we obtain
\begin{equation}
\frac{B}{\rho_{\gamma}^{1/2}}\simeq\left[1+
10\left(\frac{\lambda_{\tilde{B}}}{\lambda_H}\right)_0^2
\left(\frac{\sigma}{H}\right)_0\right]
\left(\frac{\tilde{B}}{\rho_{\gamma}^{1/2}}\right)_0\,, \label{B}
\end{equation}
where all the wavenumber indices have now been
suppressed.\footnote{It is worth noticing that at the very long
wavelength limit the effect of the gravito-magnetic interaction
described by Eqs.~(\ref{ddcB}), (\ref{rdcB}) closely resembles the
superadiabatic amplification of magnetic fields discussed
in~\cite{TW}. The most direct way of demonstrating this
resemblance is by dropping the Laplacian terms in (\ref{ddcB}) and
(\ref{rdcB}). It is then straightforward to show that the dominant
magnetic mode no longer evolves as $a^{-2}$. Instead, the field
remains constant throughout the radiation dominated epoch, while
it decreases proportionally to $a^{-1}$ during the dust (and the
reheating) era. In~\cite{TW}, superadiabatic amplification was
achieved by introducing an extra coupling between the
electromagnetic field and the extrinsic curvature of a spatially
flat FRW universe. This broke the conformal invariance of
electromagnetism on one hand, but led to a potentially huge
increase in the magnetic flux, as hyper-horizon sized fields
decayed proportionally to $a^{-1}$ instead of $a^{-2}$. Here we
see that, asymptotically, an analogous effect can be achieved
through the natural (i.e.~the purely relativistic) coupling
between the magnetic field and the Weyl curvature of a perturbed
FRW model. As we shall see in Sec.~5, the resulting increase of
the field, although not as strong as that achieved in~\cite{TW},
can still lead to a considerable boost in the overall magnetic
strength.} Expression (\ref{B}) provides the spectrum of the
``comoving'' primordial magnetic field, as it would have appeared
today were there no galactic collapse and subsequent dynamo
amplification. According to (\ref{dB}), (\ref{rB}) and (\ref{B}),
the interaction of the field with gravitational wave distortions
can change its magnitude. Crucially, the modification depends not
only on the strength of the propagating gravity waves and of the
background magnetic field, but also on the scale of the original
field. In fact, the effect on the magnitude of the magnetic field
is proportional to the square of the ratio
$(\lambda_{\tilde{B}}/\lambda_{H})_0$. Thus, superhorizon-sized
magnetic fields, like the ones considered here, interacting with
relatively strong gravity waves of comparable or even larger scale
can undergo considerable amplification.

The fact that both the perturbed and the background magnetic field
share the same (inverse-square) evolution law has two rather
important repercussions. First, it ensures that the perturbed
field does not alter the gravity wave evolution by inducing any
significant new modes (see Eq.~(\ref{ddotsigma}) and comments
immediately bellow that point). Second, it guarantees the
``legitimacy'' and continuity of Eq.~(\ref{ddotB}) throughout the
gravito-magnetic interaction. In other words, the fact that both
$B_a$ and $\tilde{B}_a$ evolve the same ensures that the system
(\ref{ddotB}), (\ref{ddotsigma}) also describes the interaction
between the gravity waves and $B_a$, once the latter has grown
strong enough to take over $\tilde{B}_a$.

\section{The application}
Hyperhorizon-sized magnetic fields emerge naturally at the end of
the inflationary era, as subhorizon quantum fluctuations in the
electromagnetic field are stretched out by the rapid expansion of
the universe. An attractive mechanism of large scale
magnetogenesis, which operates within the standard model, was
recently proposed in~\cite{D}. In their scenario, the authors
exploit the natural coupling between the Z-boson field and the
supercooled inflating gravitational background. At reheating, when
the EW-symmetry is restored, the Z-boson leads to a
(hyper)magnetic field with superhorizon correlations, which
converts into a regular magnetic field after the E-W phase
transition. The strength of the resulting field, however, is only
marginally within the galactic dynamo requirements. In particular,
the mechanism proposed in~\cite{D} produces a magnetic field of
$10^{-30}~G$, on a collapsed scale of 10~kp, when redshifted to
the epoch of galaxy formation. Although the aforementioned value
could increase by up to five orders of magnitude during reheating,
$10^{-30}~G$ is the minimum seed required for the dynamo
amplification to work, and this only if the universe is dominated
by a dark energy component (i.e.~by a cosmological constant or
quintessence)~\cite{DLT}. In addition, the value of $10^{-30}~G$
given in~\cite{D} includes the strengthening of the field, by
roughly four orders of magnitude, that occurs during the
protogalactic collapse. Next we will outline how gravity wave
distortions can amplify primordial seed magnetic fields, like
those produced in~\cite{D}, to strengths that lie comfortably
within the galactic dynamo requirements.

A collapsed magnetic field of roughly $10^{-30}~G$ with size
$\sim100$~pc, corresponds to a field of $\sim10^{-34}~G$ on a
comoving scale of about $10$~kpc~\cite{D}. The interaction of this
field with gravitational wave perturbations soon after inflation
will boost its magnitude in accordance with (\ref{B}). The
efficiency of the amplification depends on the coherence scale of
the field and on the strength of the gravitational waves.
Following~\cite{D}, we apply our analysis to a background magnetic
field with comoving strength $\tilde{B}\sim10^{-34}~G$ coherent on
a scale of $\sim10$~kpc today. The field's strength corresponds to
an energy density ratio
$\tilde{B}/\rho_{\gamma}^{1/2}\sim10^{-29}$, which remains
constant for as long as the field is frozen into the cosmic medium
and the magnetic flux is conserved. Its scale means that
$\lambda_{\tilde{B}}/\lambda_H\sim10^{20}$ at the end of
inflation, assuming that $H\sim10^{13}$~GeV~\cite{D}. At that time
the universe is also permeated by large scale gravitational waves;
the inevitable prediction of all inflationary scenarios. In fact,
the gravitational wave spectrum generated during the inflationary
expansion is perhaps the only direct signature of inflation that
might still be observable today. The energy density of these
inflation produced gravity waves is given by
\begin{equation}
\kappa\rho_{GW}\simeq{\em k}^2\left(\frac{H}{m_{Pl}}\right)^2\,,
\label{rhoGW}
\end{equation}
where ${\em k}$ is the wave's proper wavenumber and $H$ is the
Hubble parameter during inflation (e.g. see~\cite{KT}). The above
translates into the following relation for the wave induced shear
anisotropy
\begin{equation}
\Sigma_0\simeq\left(\frac{\lambda_{H}}{\lambda_{GW}}\right)_0
\left(\frac{H}{m_{Pl}}\right)\,,  \label{Sigma1}
\end{equation}
where the zero suffix indicates the end of the inflationary era.
Note that the ratio $H/m_{Pl}$ determines the vacuum energy
density of the inflaton field and, typically, the lower $H/m_{Pl}$
drops the further away from the Planck scale inflation moves.
Currently, the quadrupole anisotropy of the CMB constraints
$H/m_{Pl}$ to be less than $\sim10^{-5}$, with typical
inflationary models having $H/m_{Pl}\sim10^{-6}$. Then, on using
expression (\ref{Sigma1}), Eq.~(\ref{B}) becomes
\begin{equation}
\frac{B}{\rho_{\gamma}^{1/2}}\simeq\left[1+
10\left(\frac{\lambda_{\tilde{B}}}{\lambda_H}\right)_0
\left(\frac{\lambda_{\tilde{B}}}{\lambda_{GW}}\right)_0
\left(\frac{H}{m_{Pl}}\right)\right]
\left(\frac{\tilde{B}}{\rho_{\gamma}^{1/2}}\right)_0\,. \label{B1}
\end{equation}
Taking $\lambda_{GW}\sim\lambda_{\tilde{B}}$ initially, and
substituting the values
$(\lambda_{\tilde{B}}/\lambda_H)_0\sim10^{20}$ and
$H/m_{Pl}\sim10^{-6}$ into the above, we find that gravity wave
perturbations amplify the original field by as much as 13 or 14
orders of magnitude.\footnote{Following~\cite{D}, we are looking
at wavelengths that are already far beyond the size of the
horizon. This is the reason for the apparently ``abrupt''
amplification of the field. If, instead, we follow the mode as it
leaves the horizon and grows progressively larger, we should see a
smoother effect. Note that, as we move on towards larger and
larger scales, the magnetic boost described by Eq.~(\ref{B1})
corresponds ever more closely to an effective superadiabatic
amplification of the field analogous to that discussed
in~\cite{TW} (see also footnote in p.~7).} This means that
$B/\rho_{\gamma}^{1/2}\sim10^{-15}$, which brings the inflationary
produced seed of~\cite{D} up to $\sim10^{-21}~G$, that is
comfortably within the range of the galactic dynamo
requirements~\cite{DLT}. Note that the magnitude of
$\sim10^{-21}~G$ has been achieved without the need of any extra
amplification at reheating or during galaxy formation. Moreover,
the efficiency of the amplification also allows us to produce
magnetic fields strong enough to sustain the dynamo even in
universes with zero cosmological constant. In that case the
required value for the ratio $B/\rho_{\gamma}^{1/2}$ is raised
from $\sim10^{-25}$ to $\sim10^{-18}$ and the field itself form
$\sim10^{-30}$ to $\sim10^{-23}~G$~\cite{Ku}. Clearly, the
gravitational boost discussed here can also satisfy the latter
requirement, especially when the field's enhancement during the
protogalactic collapse is also taken into account.

\section{Discussion}
Magnetic fields have been found everywhere in the universe. Stars,
galaxies, clusters of galaxies and even high-redshift formations
carry fields that are strong and extensive. The origin of cosmic
magnetism. however, is still an unresolved problem. The nonlinear
dynamo amplification can provide galactic magnetic fields with the
desired strengths of a few $\mu G$, but requires the presence of a
seed field to start. So, where did these seeds come from? Two are
the main approaches to this question. One appeals to local
astrophysical processes, while the other advocates a primordial
origin for the seed. The attractive aspect of the latter approach
is that it can account for all the observed large scale magnetic
fields. Nevertheless, early universe magnetogenesis has its own
problems to solve. These have to do with the correlation length
and with the strength of the original seed field. Inflation can
solve the scale problem but, generally, it leads to magnetic
fields that are too weak to sustain the dynamo amplification.
Breaking the conformal invariance of the gauge fields involved is
the theoretical argument proposed as the way around the strength
problem.

Recently, an attractive inflation based mechanism, which operates
within the standard model, was suggested for the production of
large scale primordial magnetic fields~\cite{D}. The strength of
these fields is within the galactic dynamo requirements, albeit
for dark energy dominated universes and only marginally. Even when
additional amplification during the brief period of reheating is
taken into account, the achieved magnitudes are still relatively
close to the dynamo lower limits. Here, we have outlined the basic
features of a mechanism which can amplify these inflationary
produced magnetic fields to strengths that lie well within the
galactic dynamo requirements, even in conventional cosmologies
with zero dark energy component. Our approach is based on standard
general relativity and considers the interaction of large scale
primordial magnetic fields with gravitational wave distortions in
the post inflationary era. Gravity waves are inevitable byproducts
in all inflationary models, with a spectrum extending over a very
wide range of wavelengths. By allowing these waves to interact
with magnetic fields of comparable size, we found a considerable
amplification in the magnitude of the fields. As expected, the
gravitational wave boost was proportional to the induced shear
anisotropy. Crucially, however, the amplification is also
proportional to the square of the magnetic field's original
coherence length. This immediately suggested that there could be
strong gravitational amplification of magnetic fields spanning
hyperhorizon scales, like those produced in~\cite{D}. When the
actual numbers were inserted in the equations, we found that the
gravity wave induced amplification could reach up to 13 or 14
orders of magnitude. The size of the boost easily brings the
primordial fields produced in~\cite{D} up to strengths of
$10^{-21}-10^{-20}~G$, without the need of any additional
amplification at reheating or during galaxy formation. These
values are well within the dynamo limits associated with spatially
flat dark-energy dominated universes. Moreover, when the effect of
protogalactic collapse is also taken into account, the achieved
magnetic strengths lie comfortably within the dynamo requirements
associated, this time, with conventional (i.e.~zero dark-energy)
cosmologies.

\section*{Acknowledgments}
We would like to thank John Barrow, Andrew Liddle and Roy Maartens
for helpful discussions and comments. The authors also acknowledge
support from a Sida/NRF grant.


\begin{thebibliography}{99}
\bibitem[1]{K} P.P. Kronberg 1994 {\em Rep. Prog. Phys.} {\bf 57}
325; J-L Han and R. Wielebinski 2002 {\em Chin. J. Astron.
Astrophys.} {\bf 2} 293
\bibitem[2]{GR} D. Grasso and H. Rubinstein 2001 {\em Phys. Rep.}
{\bf 348} 163; L.M. Windrow 2002 {\em Rev. Mod. Pys.} {\bf 74} 775
\bibitem[3]{H} E.R. Harrison 1970 {\em M.N.R.A.S.} {\bf 147} 279;
C.J. Hogan 1983 {\em Phys. Rev. Lett.} {\bf 51} 1488; T.
Vachaspati 1991 {\em Phys. Lett. B} {\bf 265} 258; M. Gasperini,
M. Giovanini and G. Veneziano 1995 {\em Phys. Rev. Lett.} {\bf 75}
3796; G. Sigl, K. Jedamzik and A.V. Olinto 1997 {\em Phys. Rev.} D
{\bf 55}, 4582; M. Joyce and M. Shaposhnikov 1997 {\em Phys. Rev.
Lett.} {\bf 79} 1193; E. Calzetta and A. Kandus 2002 {\em Phys.
Rev. D} {\bf 65} 063004
\bibitem[4]{D} A-C Davis, K. Dimopoulos, T. Prokopec and O.
T\"{o}rnkvist 2001 {\em Phys. Lett. B} {\bf 501} 165; K.
Dimopoulos, T. Prokopec, O. T\"{o}rnkvist and A-C. Davis 2002 {\em
Phys. Rev. D} {\bf 65} 063505
\bibitem[5]{P} E.N. Parker {\em Cosmical Magnetic Fields}
(Clarendon, Oxford 1979); Y.B. Zeldovich, A.A. Ruzmaikin and D.D.
Sokoloff {\em Magnetic Fields in Astrophysics} (McGraw-Hill, New
York 1983); P.P. Kronberg 1994 {\em Rep. Prog. Phys.} {\bf 57}
325; R. Beck, A. Brandenburg, D. Moss, A.A. Shukurov and D.
Sokoloff 1996 {\em Annu. Rev. Astron. Astrophys.} {\bf 34} 155
\bibitem[6]{Ku} R.M. Kulsrud, in {\it Galactic and Extragalactic
Magnetic Fields}, Eds R. Beck, P.P. Kronberg, and R. Wielebinski
(Reidel, Dordrecht, 1990).
\bibitem[7]{DLT} A.C. Davis, M. Lilley and O. T\"{o}rnkvist 1999
{\em Phys. Rev. D} {\bf 60} 021301
\bibitem[8]{KA} R.M. Kulsrud and S.W. Anderson 1992 {\em
Astrophys. J.} {\bf 396} 606
\bibitem[9]{C} J.M. Cornwall 1997 {\em Phys. Rev. D} {\bf 56}
6146; D.T. Son 1999 {\em Phys. Rev. D} {\bf 59} 063008; G.B. Field
and S.M. Carroll 2000 {\em Phys. Rev. D} {\bf 62} 103008; M.
Christensson, M. Hindmarsh and A. Brandenburg 2001 {\em Phys. Rev.
E} {\bf 64} 056405
\bibitem[10]{Do} A.D. Dolgov 1981 {\em Sov. Phys. JETP} {\bf 54}
223; A.D. Dolgov 1993 {\em Phys. Rev. D} {\bf 48} 2499
\bibitem[11]{TW} M.S. Turner and L.M. Windrow 1988 {\em Phys. Rev.
D} {\bf 30} 2743
\bibitem[12]{CKM} E.A. Calzetta, A. Kandus and F.D. Mazzitelli
1998 {\em Phys. Rev. D} {\bf 57} 7139; A. Kandus, E.A. Calzetta,
F.D. Mazzitelli and C.E.M. Wagner 2000 {\em Phys. Lett. B} {\bf
472} 287; B.A. Bassett, G. Pollifrone, S. Tsujikawa and F.
Viniegra 2001 {Phys. Rev. D} {\bf 63} 103515
\bibitem[13]{GS} M. Giovannini and M. Shaposhnikov 2000 {\em Phys.
Rev. D} {\bf 62} 103512
\bibitem[14]{BM} O. Bertolami and D.F. Mota 1999 {\em Phys. Lett.
B} {\bf 455} 96; A. Mazumdar, M.M. Sheikh-Jabbari 2001 {\em Phys.
Rev. Lett.} {\bf 87} 011301; M. Giovannini 2000 {\em Phys. Rev. D}
{\bf 62} 123505
\bibitem[15]{E} G.F.R. Ellis, in {\em Carg\`{e}se Lectures in
Physics}, vol. 1, edited by E. Schatzman (Gordon and Breach, New
York, 1973) p. 1; G. F. R. Ellis and H. van Elst, in {\it
Theoretical and Observational Cosmology}, edited by M.
Lachi\`eze-Rey (Dordrecht: Kluwer 1999) p. 1
\bibitem[16]{TB} C.G. Tsagas and J.B. Barrow 1997 {\em Class.
Quantum Grav.} {\bf 14} 2539; C.G. Tsagas and J.D. Barrow 1998
{\em Class. Quantum Grav.} {\bf 15} 3523; C.G. Tsagas and R.
Maartens 2000 {\em Phys. Rev. D} {\bf 61} 083519; C.G. Tsagas and
R. Maartens 2000 {\em Class. Quantum Grav.} {\bf 17} 2215
\bibitem[17]{DBE} P.K.S. Dunsby, B.A.C.C. Bassett and G.F.R. Ellis
1997 {\em Class. Quantum Grav.} {\bf 14} 1215; A. Challinor 2000
{\em Class. Quantum Grav.} {\bf 17} 871
\bibitem[18]{MTU} R. Maartens, C.G.Tsagas and C. Ungarelli 2001
{\em Phys. Rev. D} {\bf 63} 123507; C.G. Tsagas 2002 {\em Class.
Quantum Grav.} {\bf 19} 3709
\bibitem[19]{MDB} M. Marklund, P. K. S. Dunsby and G. Brodin 2000
{\em Phys. Rev. D} {\bf 62} 104008
\bibitem[20]{KT} E.W. Kolb and M.S. Turner {\em The Early
Universe} (Addison-Wesley 1990); T. Padmanabhan {\em Structure
Formation in the Universe} (Cambridge University Press, Cambridge
1993); J.A. Peacock {\em Cosmological Physics} (Cambridge
University Press, Cambridge 1999); A.R. Liddle and D.H. Lyth {\em
Cosmological Inflation and Large-Scale Structure} (Cambridge
University Press, Cambridge 2000)
\end{thebibliography}
\end{document}